\documentstyle[11pt,aas2pp4]{article}

\begin{document}

\pagestyle{empty}

\hspace{0mm}\\[20mm]
{\large {\bf RADIO PULSAR TIMING}} 
\vskip 1.0truecm
J. F. Bell
\vskip 1.0truecm
{\it The University of Manchester, NRAL, Jodrell Bank, Macclesfield, Cheshire
SK11~9DL, UK.}
\vskip 1.5truecm

ABSTRACT \vskip 0.5truecm

The motivation for radio pulsar timing and its basic principles are
reviewed.  Present and future radio timing techniques and hardware are
summarised and compared. The array of present timing programmes and their
scientific goals are collated and described. Recent results and future
prospects are discussed, with emphasis on multi-wavelength techniques where
appropriate. Timing of radio pulsars at other wavelengths is summarised
along with the provision of contemporary ephemerides for timing and searches
at other wavelengths.\vskip 0.5truecm

INTRODUCTION: MOTIVATION FOR TIMING\vskip 0.5truecm

The scope of this paper is confined to timing of rotation powered radio
pulsars, as there are many papers in session E1.1 of the COSPAR meeting
discussing the timing of accretion powered X-ray pulsars. For this paper,
timing will be considered to be semi-regular post-detection measurements of
arrival times. In the following two sections, the motivation for radio
pulsar timing is discussed in terms of understanding pulsars themselves as
well as their use as tools.\vskip 0.5truecm

%Conducting a survey of the papers
%that have the NASA/STI keyword ``pulsars'' in the ADS
%(http://adsabs.harvard.edu/abstract\_service.html) abstract database gives
%4546 papers between 01/01/1974 and 31/12/1994. The distribution with time is
%surprisingly flat with a mean of 227 papers per year and a standard
%deviation of 24. On closer inspection is becomes apparent that there is a
%strong two year periodicity in the number of pulsar papers published per
%year. Summing over the odd numbered years gives 2471 papers while summing
%over the even numbered years gives 2076 papers. \vskip 0.5truecm

\underline{Neutron Star Properties, Formation and Evolution} 

\begin {itemize}

\item {\it Neutron Star Structure:} Pulsars provide an excellent opportunity
to study the structure of neutron stars as Alpar has described at this
meeting. The steady spin down of some pulsars is interrupted by occasional,
sudden increases of rotation rate. These events, known as glitches, have
provided strong evidence that neutron stars consist of a solid but brittle
crust and a fluid interior. This fluid interior is generally identified with
a neutron superfluid. The origins of glitches have been identified with
adjustments in ellipticity of the neutron star as it spins down \cite{rud69}
and/or an irregular outflow of angular momentum due to the quantized vortex
properties of the interior neutron superfluid \cite{ai75}.

\item {\it Equations of State:} The properties of neutron stars that can be
determined observationally in order to constrain the equation of state are
the mass, radius and limiting spin period \cite{vp96}. Timing measurements
shed light on these properties in a number of ways. Measurements of Shapiro
delay and other relativistic effects in some cases give very accurate mass
determinations \cite{ktr94}. Masses may also be constrained using radial
velocity curves of both pulsars and their optical companions \cite{vbk96}.
In conjunction with ages determined from white dwarf cooling, rotation
periods of the fastest millisecond pulsars (MSPs) at birth may be
constrained \cite{kul92,bkb+95}. The next generation of surveys (which will
be sensitive to periods less than one millisecond) will provide constraints
as described by D'Amico in these proceedings.

\item {\it Pulsar Velocities:} The mean birth velocity of normal
pulsars (some of which were measured from timing) is 450~km~s$^{-1}$
\cite{ll94}. These large velocities of normal pulsars have been taken as firm
evidence for velocity kicks from asymmetric supernovae. Direct evidence for
a substantial birth kick has recently been observed in the precessing binary
PSR~J0045$-$7319 \cite{kbm+96}.

\item {\it Stellar Evolution:} There are several different models for the
formation of MSPs and at this stage none can be ruled out \cite{pk94}.
Timing is providing clues in unravelling formation history, in terms
of velocities \cite{cmb+97,nt95}, accurate positions for studying optical
companions \cite{kul86,bbb93,lfc96,van96}, and eccentricities which provide
a useful test of evolutionary models \cite{phi92b}. Timing of eclipsing
millisecond pulsars has provided some very important insights into the
nature of their companions and their evolution \cite{sbl+96,nt96}.

\end{itemize}

\underline{Pulsars as Tools}\vskip 0.5truecm

Radio pulsar timing is a field that is interesting not only to study the
objects themselves, but also because it is a valuable tool for fundamental
tests of physics and many other applications.

\begin {itemize}

\item {\it Tests of Relativity:} The realisation that radio pulsars are
superb natural laboratories for testing theories of gravity came with a
flurry of publications after the discovery of PSR B1913+16 \cite{ht74}.
Measurement of several relativistic effects allowed a complete solution of
the system's parameters and a self-consistent confirmation of the rate of
emission of gravitational waves as predicted by general relativity
\cite{tw89}. Recently, the fifth relativistic dual-neutron-star binary
pulsars was been found \cite{nst96}. When analysed together, these provide
even more stringent tests of theories of gravity \cite{twdw92}. These
pulsars have also been used to set limits on variations of the gravitational
constant $G$. The best limit to date is from timing of PSR B1913+16 which
gives $\dot{G} /G = 4 \pm 5
\times 10^{-12}$~yr$^{-1}$ \cite{tay92}, although the limit from PSR~B1855+09
should improve on this in the future \cite{ktr94}. A similar result has been
obtained by assuming that present masses of neutron stars reflect the
Chandrasekhar mass and thus the value of G when they were formed
\cite{tho96}. There is also a range of interesting and useful tests that can
be performed using neutron stars in very circular orbits ($e \sim 1 \times
10^{-5}$): tests of the strong equivalence principle \cite{ds91,wex96}, of
local Lorentz invariance \cite{de92a} and of conservation laws
\cite{bd96}. These latter tests depend linearly on the orbital eccentricity
of a sample of binary millisecond pulsars.

\item {\it Interstellar Medium:} Pulsars are excellent tools for studying
the interstellar medium (ISM) because they are pulsed point sources. The
radio dispersion depends on the column density of free electrons between the
pulsar and the observer. Hence measurements of dispersion yield free
electron column density estimates for those pulsars which have independently
determined distances \cite{lmt85,tc93}. Since the magnetic fields in the ISM
lead to Faraday rotation, polarisation studies of linearly polarised pulsars
can give estimates of ISM magnetic field strengths \cite{hl88}. Scattering
on smaller scales leads to rapid variations in the observed flux, called
scintillations \cite{ric90}. Dispersion measure variation
\cite{bhvf93,ktr94} and diffractive scattering \cite{cpl86} observations
provide constraints on the scales of density fluctuations in the ISM and can
also be used to determine pulsar velocities \cite{gup95}.

\item {\it Time Standards:} Very precise long term timing of PSRs B1937+21
and B1855+09 have demonstrated the long term clock-like stability of MSPs
\cite{ktr94}. Based on this stability, there have been suggestions that MSPs
could be used as a new standard of time. Unfortunately, a standard of time
based on pulsars would have no link to a reproducible physical phenomenon,
whereas the atomic clocks do \cite{bau90}. Nevertheless, pulsars such as PSRs
B1855+09 and J1713+0747 appear to be more stable on time scales of months to
years than atomic clocks. Hence, with enough MSPs well distributed over the
celestial sphere, it might be possible to use an ensemble-averaged pulsar
time scale as a long term time standard.

\item {\it Reference Frame Ties:} The tying of the radio, optical and
planetary reference frames is a difficult proposition. It requires large
numbers of randomly distributed sources with accurately determined positions
and proper motions in all of the three reference frames. Pulsars provide a
good link between the planetary and radio frames as accurate positions can
be obtained in each frame.  Binary millisecond pulsars are helpful as many
have positions and proper motions that will be measurable in all three
frames \cite{bbm+95,lfc96}.  Several VLBI and optical programmes are
presently obtaining observations towards this end.

\item {\it Cosmology:} Another novel use of pulsars as tools has been to
place limits on the cosmic background of low frequency gravitational waves
\cite{rt83}. The best limit is $\Omega_{g} h^{2} < 6 \times 10^{-8}$, from
long term timing data on PSR B1855+09, where $\Omega_{g}$ is the fractional
energy density in gravitational waves per logarithmic frequency interval and
the Hubble constant $H_{0} = 100 h$~km~s$^{-1}$ \cite{td96,ktr94}. Since
such a stochastic background would be produced by cosmic strings, this
suggests that cosmic string are rather unlikely to exist \cite{hr84}.

\item {\it Search for Planets:} The discovery of planetary mass companions
to a millisecond pulsar \cite{wf92} was unexpected.  Their existence has been
confirmed by the detection of three-body perturbations in the timing
residuals \cite{wol94}.  Naturally, such a discovery creates interest in
finding other MSPs with planetary mass companions. Several efforts to search
for planets around other pulsars \cite{tp92,bbm+96,she95a} have provided only
a couple of candidates to date \cite{bls93,bfs93}.

\item {\it Solar System Ephemerides:} An accurate ephemeris for the major
bodies of the solar system \cite{sta90} is vital for transforming
topocentric pulse arrival times to barycentric pulse arrival times. The most
precisely timed pulsars offer an independent test bed as correlated
irregularities in timing data can provide feedback, allowing different
ephemerides to be compared and improved \cite{kas95}.

\end{itemize} \vskip 0.2truecm

TIMING PRINCIPLES\vskip 0.5truecm

\underline{Measuring Arrival Times}\vskip 0.5truecm

The time-of-arrival (TOA) of a fiducial point in the rotational phase of a
pulsar is the fundamental quantity which must be determined. This is
normally done by comparison with a standard pulse profile $s(t)$. The
observed pulse profile $p(t)$ can be expressed in terms of the standard
profile by $p(t) = a + b s(t - \phi) + g(t)$ where $a$ is a DC offset, $b$
is a scale factor, $\phi$ is a phase shift and $g(t)$ represents noise. For
the comparison, full use of the available signal to noise is most easily
achieved by cross-correlating the observed and standard profiles in the
Fourier domain \cite{tay92}. To do this a very accurate time standard is
required and is usually obtained from a local hydrogen maser referenced to a
standard bank of caesium clocks \cite{bau90}. \vskip 0.5truecm

\underline{Transformation to the Pulsar Rest Frame} \vskip 0.5truecm

Having determined the topocentric TOA $t$, an accurate transformation to the
rest frame of the pulsar must be applied. For most single pulsars, a
transformation to the barycentre of the solar system is sufficient
\cite{bh86,tw89}. The barycentric time of arrival is
\begin{equation}
\label{e:bary}
t_{b} = t + {{\bf r \cdot \hat{n}} \over c} + 
{({\bf r \cdot \hat{n}})^2 - |{\bf r}|^2 \over 2 c d }
- {D \over f^2 } + \Delta_{E \odot} + \Delta_{S \odot} + \Delta_{A \odot},
\end{equation}
where ${\bf r}$ is a vector from the barycentre to the phase centre of the
telescope, ${\bf \hat{n}}$ is a unit vector pointing from the barycentre to
the pulsar, $c$ is the speed of light, $d$ is the distance to the pulsar,
$D$ is the interstellar dispersion constant, $f$ is the radio frequency,
$\Delta_{E \odot}$ is the Einstein delay comprised of the gravitational red
shift and time dilation, $\Delta_{S \odot}$ is the Shapiro delay
characterising the curvature of space time near the sun and $\Delta_{A}$ is
the aberration delay as a result of the Earth's rotation \cite{tw89}. Terms
two and three together make up the Roemer delay $\Delta_{R\odot}$. The first
part of the 3rd term measures the curvature of the wavefronts emitted from
the pulsar and can be used to determine a timing parallax for nearby
pulsars. The dispersion measure and five astrometric (right ascension,
declination, parallax, proper motion in right ascension and declination)
pulsar parameters can be determined via this transformation. \vskip
0.5truecm

\underline{High Frequency Timing.} At higher frequencies such as infrared,
optical, X-ray and gamma rays, dispersive effects are no longer important and
$D/f^{2} \longrightarrow 0$. For ground based optical and infrared
observations, the data analysis is therefore considerably simpler. For space
or balloon based observations extra transformations (including special
relativistic effects) are required to account for the more complicated motion
of the observatory. Further, it is not always possible to have a sufficiently
accurate clock at the telescope, requiring regular determination of clock
offsets.\vskip 0.5truecm

\underline{Binary Pulsars.} Further terms are needed to account for binary
motion and to obtain the time $T$ in an inertial frame with respect to
the pulsar's centre of mass. This transformation is given by
\begin{equation}
\label{e:pbary}
t_{b} = T + \Delta_{R} + \Delta_{E} + \Delta_{S} + \Delta_{A} 
\end{equation}
where $\Delta_{R}$, $\Delta_{E}$ and $\Delta_{S}$ are the Roemer, Einstein
and Shapiro delays in the binary orbit and $\Delta_{A}$ is the aberration
delay as a result of the pulsar's rotation \cite{tw89}. The five Keplerian
parameters necessary to describe the orbit and account for the Roemer delay
are: orbital period, orbital eccentricity, longitude of periastron, epoch of
periastron and the projected semi-major axis. A further eight post-Keplerian
parameters might be measurable and account for the final three terms in
equation \ref{e:pbary} \cite{dt92}. In those rare but exciting cases where a
pulsar has more than one companion an extra complete set of the four terms
$\Delta_{R}$, $\Delta_{E}$, $\Delta_{S}$ and $\Delta_{A}$ may be required
for each orbit. The above only considers companions that are sufficiently
compact to act like point masses, for example neutron stars, white dwarfs
and planets. For pulsars with larger or less dense companions, such as
PSR~J0045$-$7319 and possibly PSRs~J2051$-$0827 and B1259$-$63 additional
terms are required to account for the quadrupole moment \cite{lbk95}. \vskip
0.5truecm

\underline{Pulsar Spindown and Dynamics}\vskip 0.5truecm

Having made the above transformations, the pulsar's rotational
parameters are now accessible by measuring the rotational phase
$\phi(t)$ given by a Taylor expansion
\begin{equation}
\label{e:phase}
\phi(t) = \phi(0) + \nu t + {1 \over 2} \dot{\nu}t^{2} + {1 \over 6}
\ddot{\nu} t^{3} + .......
\end{equation}
where $\nu \equiv 1/P $ is the rotational frequency, and
$\dot{\nu},\ddot{\nu}$ are the frequency derivatives corresponding to the
period derivatives $\dot{P},\ddot{P}$. The rotational parameters include a
fiducial time defining the phase, the rotation period, and possibly several
period derivatives. The values these parameters take are those measured by an
observer at infinite distance. To obtain the actual values one must correct
for the gravitational redshift of the pulsar. This would typically reduce the
value for the period by 30\%, however this may vary by a factor of two
depending on the equation of state of the neutron star \cite{cst94}. Since
the many theoretical equations of state for neutron stars are poorly
constrained by observations, such corrections are not normally applied.\vskip
0.5truecm

As a pulsar may be moving toward or away from the Earth, its measured
period will include both the actual period and a contribution from the
Doppler effect. While this effect cannot be separately measured, the
difference between the measured period and actual period is only 0.1\% for a
pulsar having a radial velocity of 300~km~s$^{-1}$. The orbital periods of
binary pulsars are similarly affected. Pulsars may also have accelerations
towards or away from us and the measured period derivatives and orbital
period derivatives contain a contribution from the Doppler effect
\cite{shk70,dt91,ctk94,bb96}. Transverse motions can be measured separately
from timing \cite{mtv74} and indeed transverse velocities for many pulsars
have been obtained in this way. While this technique has proved to be
difficult for young pulsars due to timing noise \cite{ls90}, more recently
the precision with which MSPs can be timed has facilitated proper motion
measurements for about half the Galactic population of MSPs
\cite{cmb+97}. Recently the effect of proper motion on the observed
semi-major axes of binary pulsars has been shown to be important for fast
pulsars in wide face-on binaries \cite{ajrt96,kop96}.\vskip 0.5truecm

\newpage

\underline{Dispersion Removal}\vskip 0.5truecm

In the pursuit of the highest possible timing precision, the limiting factor
is may often be either signal-to-noise or the dispersion smearing of the
pulses for those pulsars that do not have significant timing noise
\cite{tay96}.  The need for improved sensitivity inevitably leads to
bandwidths as large as possible being used.  As a result, the dispersion of
the pulses in time across this bandwidth at best leads to broadening of the
observed pulse and for fast pulsars, may amount to many pulse
periods. Removing the effects of dispersion is therefore of fundamental
importance for precise timing and the accuracy with which this can be done
limits many of the presently running pulsar timing projects. The properties,
advantages and disadvantages of various methods of attacking this problem
are summarised in the following list.

\begin{itemize}

\item {\it Filter Banks} offer a relatively cheap and reliable way of
removing dispersion and have been used in numerous very functional,
productive and scientifically potent timing programmes. In recent times, the
the need for very narrow channel bandwidths as well as stable and
well-matched gains has exposed the limitations of filter banks for precise
timing of MSPs.

\item {\it Correlators} provide the means to overcome these limitations, in
particular the gains of the ``lags'' can be calibrated, giving more timing
stability \cite{nav94}. However, correlators are substantially more costly
and complex to build.

\item {\it Frequency Sweeping} provides a useful method for dispersion
removal and is based on a swept local oscillator which closely follows the
frequency drift of the pulsar signal \cite{mtw79,bab+89}. To date this has
not been a widely explored technique and its inflexibility and hardware
limitations have made it possible only for intermediate dispersion measures
of around 70\,cm$^{-3}$pc at 1400 MHz. However some very precise timing has
been achieved \cite{cbl+93,cbl+95}.

\item {\it Coherent Dedispersion,} in which the raw base-band signal is
recorded prior to square-law detection, after which software transforms are
applied to remove dispersion, offers several advantages if the recording
system has enough bandwidth to allow sufficiently fast sampling
\cite{han71}. The frequency resolution is simply a software parameter so
that as new fast computers arrive, the accuracy of the timing experiment can
be steadily improved while the costly in-house built hardware of the above
methods becomes obsolete \cite{tay96,hsr87,ajk+96}.  Although these ideas
have been around for a while, only in the last couple of years have the tape
and computing resources become available to make this method feasible for
substantial timing projects.

\item {\it Multi-frequency Timing} has demonstrated that dispersion measure
variations are one of the more important factors limiting timing precision
\cite{bhvf93,ktr94}. The path which most groups are now following or are
intending to follow is that of simultaneous timing at two or more
frequencies. This allows the dispersion measure variations to be determined
so that appropriate corrections to timing parameters can be made
\cite{ktr94}.

\end{itemize}
\newpage

TIMING PROGRAMMES \vskip 0.5truecm

In this section some of the large and long term timing programmes and their
recent results are summarised. A more complete table of timing programmes,
principle contacts, sources lists and observation intervals can be found on
the WWW (http://astro.berkeley.edu/$\sim$mpulsar/).\vskip 0.5truecm

\underline{Millisecond Pulsars}\vskip 0.5truecm

Using Arecibo, the Princeton group has set the standard of precision
timing, starting with the millisecond pulsars B1937+21, B1855+09
\cite{ktr94} and also J1713+0747 \cite{cfw94}. At 1400 MHz, the rms
residuals for two of these pulsars are at the 0.3\,$\mu$s level. There are
now 14 MSPs in the timing list, observed at roughly fortnightly intervals
(except while the present Arecibo upgrade is taking place). Eight other MSPs
are also being timed at the VLA by the Princeton group.  The Penn State
group, also using Arecibo, have confirmed the very exciting planets around
PSR~B1257+12 \cite{wol94}. Using the 42m and 25m at Green bank, the Berkeley
group has conducted multi-frequency timing of 15 MSPs, some for up to six
years, and has produced some useful results including detailed studies of
dispersion measure variations \cite{bhvf93,bw96} and independent
confirmation of the planets around PSR~B1257+12 \cite{bsf92}. \vskip
0.5truecm

The Nan\c{c}ay group has achieved some impressive results, timing
PSR~B1821$-$24 and PSR~B1937+21 since 1988, achieving rms timing residuals
of 0.3\,$\mu$s for PSR B1937+21 \cite{cbl+93,cbl+95}.  The Parkes and
Jodrell groups have been timing all MSPs including those in globular
clusters, since their discovery. The highest levels of precision have not
yet been achieved since to a large degree the observed profiles are
dominated by dispersion smearing \cite{bbm+96,cmb+97}. The rms residuals are
steadily improving, particularly with the vastly improved frequency
resolution afforded by the Caltech correlator \cite{nav94,sbm+96} at Parkes
and the new timing machine at Jodrell. Some interesting results have been
obtained with high frequency timing at Effelsberg, including pulse shape
variations of PSR J1022+1001 \cite{cnk+97}. \vskip 0.5truecm

\underline{High Mass Binary Pulsars}\vskip 0.5truecm

The two groups of pulsars categorised as high mass binaries are the
dual-neutron-star binaries such as PSR~B1913+16 and those with high mass
young stellar companions, PSR~B1259$-$63 and PSR~J0045$-$7319. The known
dual-neutron-star binaries are all in the northern hemisphere and the long
term timing programme of the Princeton group has produced the most results,
including confirmation of the predictions of general relativity at better
than the 1\% level \cite{dt91,twdw92}. At its last periastron passage in
January 1994, PSR~B1259$-$63 revealed some remarkable interactions with its
Be star companion, including the disappearance of the pulsar, huge
dispersion measure and rotation measure variations and a continuum source
possibly due to a wind interaction or jet \cite{jml+96,joh96}. Recent X-ray
observations \cite{ktn+95,khnt96} are more easily explained by the wind
interaction model \cite{tak94}. PSR~J0045$-$7319 is a binary system with a B
star and shows no frequency dependent variations of any kind
\cite{ktm96}. However the orbital plane has recently been
shown to be inclined to the rotational axis of the B star, giving rise to
spin-orbit coupling and precession of the pulsar orbit \cite{kbm+96}. This
has provided independent evidence for a substantial neutron star birth
kick. \vskip 0.5truecm

\underline{Glitching Pulsars}\vskip 0.5truecm

Long-term regular monitoring of old favourites including the Crab and Vela
has been continued by most groups and several new glitches have been
observed \cite{lps93,arz95} including a giant glitch in PSR~B1757$-$24
\cite{lkb+96}. Some groups continue to monitor individual pulsars such as
Vela for up to 18 hours per day \cite{mcc96}. A large number of young pulsar
were found in high-frequency Galactic plane surveys \cite{cl86,jlm+92} and
the timing of these and other pulsars has resulted in the discovery of over
25 glitches \cite{sl96}. This large number of glitches, now 50 in total, has
allowed statistical studies which indicate that post-glitch relaxations
can be separated into two exponential components for most pulsars
\cite{sls96}. Alpar (this meeting) has shown that the majority of
glitches are inconsistent with the crust cracking model, but are explained
by the superfluid vortex pinning model. Recently the long time span of data
on Vela has been collated and used to demonstrate that its braking index is
$1.4 \pm 0.2$ \cite{lpgc96} which is impressive given the frequency and
magnitude of its glitches. \vskip 1.2truecm

\underline{Timing at Other Wavelengths}\vskip 0.5truecm

Radio pulsar timing at wavelengths other than radio until recent times has
been mostly restricted to two pulsars, the Crab and PSR~B0540$-$69 at
optical wavelengths, although there was also some timing of Vela
\cite{mwpe80}. Caraveo and others (this proceedings) give more detailed
discussions of timing at these wavelengths. There was extensive optical
timing of the Crab pulsar during the 1970's \cite{loh81} which resulted in
the detection of glitches \cite{loh75,gro75c} and also studies of the random
walk nature of timing noise \cite{cor80}. Recently there have been
simultaneous timing studies of the radio giant pulses and gamma-ray
pulsations \cite{lcu+95}. PSR~B0540$-$69 was discovered in X-rays
\cite{shh84} but most of the timing, including measurement of the
braking index has been at optical wavelengths
\cite{mpb87,mp89,gfo92,bcd+95}. It was eventually detected in the radio
after some very long integrations by radio search standards
\cite{mml+93}. Recently braking indices for the Crab and PSRs B0540$-$69 and
B1509$-$58 have been determined from X-ray timing observations using Ginga
\cite{ndl+90}. The gamma-ray pulsar Geminga has not been detected at radio
wavelengths but is mentioned here due to its similarity to Vela and other
radio pulsars. There has been a considerable amount of timing of Geminga, in
particular using COS B \cite{bc92,gbb+94} and SAS 2 \cite{mbf92} which
accumulated over 10 years of timing data. \vskip 0.5truecm

\underline{Contemporary Ephemerides.} A fundamental reason for the provision
of contemporary ephemerides for timing and searches at other wavelengths is
that in many cases, less than one photon per pulse period is observed. For
example the average separation between photons from the Crab when observed
by EGRET is 10 minutes, which corresponds to about 18000 pulse periods
\cite{tho93}. The resulting long integration times may lead to pulse
smearing if period derivatives and binary motion are not accounted
for. Another important reason is that a 3 sigma event is only a detection if
the search parameter space is small and if the noise is normally
distributed. If a range of frequencies in a power spectrum must be searched
then the significance level of a peak of power $P$ is $S = (1 - e^{-P})^{N}$,
where $N$ is the number of frequencies searched \cite{ptvf92}. If $N$ is
large, $S$ very quickly becomes a small number, unless $P$ is also
large. Further, noise in a power spectrum is exponentially distributed, not
normally distributed. However, if the data can be folded according to an
accurate ephemeris, much weaker signals can be reliably detected.\vskip
0.5truecm

\underline{Present Arrangements.} Previously, most of the distribution of
up-to-date ephemerides has been by personal contact, often via the Princeton
email exploder (psrtime@pulsar.princeton.edu). There have of course been
some pulsars for which more organised processes have been established such
as the GRO database provided by Jodrell Bank, Parkes and Princeton which is
maintained at Princeton (ftp pulsar.princeton.edu). Another example is the
Crab pulsar ephemerides \cite{lps93} which have been available for the past
12 years and have been on the WWW (http://www.jb.man.ac.uk/$\sim$pulsar/)
for a while. \vskip 0.5truecm

\underline{Future Possibilities.} The Princeton group provide
a WWW search facility (http://pulsar.princeton.edu/) through which pulsar
parameters can be readily obtained. This server has provided a good
prototype, offering rapid, simple access to many pulsars. The Jodrell Bank
group will shortly be providing a similar but vastly more extensive
database, containing up-to-date ephemerides for every pulsar observable from
Jodrell, ie, parameters for over 300 pulsars
(http://www.jb.man.ac.uk/$\sim$pulsar/).  Hopefully other groups with
regular timing programmes will provide similar access in the future. \vskip
0.5truecm

FUTURE PROSPECTS \vskip 0.5truecm

In this section, some of the areas in which substantial progress is likely
to be made in the next few years are noted. A large fraction of these relate
to MSPs, as most of them have only recently been discovered and many new
effects are now accessible due to the high timing precision
attainable. \vskip 0.5truecm

\underline{Timing Noise and Time Standards} \vskip 0.5truecm

An enormous timing database (equivalent to over 3000 years of data on a
single pulsar) has been collected at Jodrell Bank and is presently being
analysed \cite{mlp97}. This will allow a very complete understanding of
timing noise and improve on previous studies \cite{arz95}. The time span of
data collected for a number of MSPs will soon be long enough to assess the
extent to which timing noise also occurs for the fastest pulsars
\cite{tay96}. In this respect the next few years should establish whether or
not pulsars can really be used as long term standards of time.\vskip
0.5truecm

\underline{Dynamical Effects} \vskip 0.5truecm

It should be possible to improve distance estimates to MSPs not only using
interferometric techniques, but also by measuring timing parallaxes
\cite{rt91a,cfw94,sbm+96} and orbital period derivatives \cite{bb96}.  If
accurate distances to some high Galactic latitude binary MSPs can be
obtained, it may be possible to constrain dark matter in the local Galactic
disk \cite{bb96}. Many more proper motions should be measurable
\cite{cmb+97} and when combined with accurate distances, will allow
intrinsic period derivatives to be determined \cite{ctk94,shk70}. Useful
constraints on orbital inclinations may come from measurements of the proper
motion contribution to the apparent semi-major axis \cite{kop96}, while
pulsar orbital parallaxes \cite{kop95} may require more substantial improves
in timing precision. As the timing residuals reduce for many of the more
recently discovered MSPs, there are likely to be a few more for which Shapiro
delay can be measured, providing constraints on companion masses. \vskip
0.5truecm

\underline{Tests of Relativity} \vskip 0.5truecm

Important progress is likely to be made in constraining theories of gravity
by combining the results from several dual-neutron star binaries
\cite{twdw92,dt92}, although nearby objects will need interferometric
distances \cite{bb96} if they are to live up to their full potential
\cite{arz95}. For the other tests and for limits on $\dot{G}$ and
ultra-low-frequency gravitational waves (mentioned in the introduction),
small improvements will come with continued observations. Substantial
improvements will require the discovery of new more extreme systems, novel
methods and new effects. There are several gravitational lensing
effects that occur if a binary pulsar orbit is sufficiently edge on
\cite{sch89b,sch90,dk95} that are presently extremely difficult to
measure. \vskip 0.5truecm

\underline{Speculation} \vskip 0.5truecm

If the polarisation of MSPs can be understood, it may be possible to
determine the inclination of the spin axes and therefore the orbits (if the
angular momenta are assumed to be aligned). While the lack of short orbital
period planets around MSPs seems to be established at this time
\cite{bbm+96}, the existence of planets like those in the outer solar system
will be tested over the next few years. The next generation of telescopes such
as a square kilometer array offer many exciting prospects for pulsar
timing. The discovery of novel effects and new types of systems such as black
hole pulsar binaries is also possible. \vskip 0.5truecm

ACKNOWLEDGEMENTS \vskip 0.5truecm

It is with pleasure that I thank A. Lyne, V. Kaspi and F. Camilo for
stimulating discussions and constructive comments on this paper. \vskip
0.5truecm

\small

%\bibliographystyle{apj1d}
%\bibliography{journals1,modrefs,psrrefs}

\begin{thebibliography}{}

\bibitem[Anderson \& Itoh 1975]{ai75}
Anderson, P.~W. \& Itoh, N. 1975, Nature, 256, 25

\bibitem[Anderson {et al.}  1996]{ajk+96}
Anderson, S.~B., Jenet, F.~A., Kaspi, V.~M., Prince, T.~A., Sandhu, J.~S.,
  Unwin, S.~C., \& Navarro, J. 1996, in { Pulsars: Problems and Progress, {IAU}
  Colloquium 160}, ed.\ S. Johnston, M.~A. Walker, \& M. Bailes, Astronomical
  Society of the Pacific, 211

\bibitem[Arzoumanian 1995]{arz95}
Arzoumanian, Z. 1995.
\newblock PhD thesis, Princeton University

\bibitem[Arzoumanian {et al.}  1996]{ajrt96}
Arzoumanian, Z., Joshi, K., Rasio, F., \& Thorsett, S.~E. 1996, in { Pulsars:
  Problems and Progress, {IAU} Colloquium 160}, ed.\ S. Johnston, M.~A. Walker,
  \& M. Bailes, Astronomical Society of the Pacific, 525

\bibitem[Backer, Sallmen, \& Foster 1992]{bsf92}
Backer, D., Sallmen, S., \& Foster, R. 1992, Nature, 358, 24

\bibitem[Backer, Foster, \& Sallmen 1993]{bfs93}
Backer, D.~C., Foster, R.~F., \& Sallmen, S. 1993, Nature, 365, 817

\bibitem[Backer {et al.}  1993]{bhvf93}
Backer, D.~C., Hama, S., Van~Hook, S., \& Foster, R.~S. 1993, ApJ, 404, 636

\bibitem[Backer \& Hellings 1986]{bh86}
Backer, D.~C. \& Hellings, R.~W. 1986, ARAA, 24, 537

\bibitem[Backer \& Wong 1996]{bw96}
Backer, D.~C. \& Wong, T. 1996, in { Pulsars: Problems and Progress, {IAU}
  Colloquium 160}, ed.\ S. Johnston, M.~A. Walker, \& M. Bailes, Astronomical
  Society of the Pacific, 87

\bibitem[Bailes, Lyne, \& Shemar 1993]{bls93}
Bailes, M., Lyne, A.~G., \& Shemar, S.~L. 1993, in { Planets Around Pulsars},
  ed.\ J.~A. Phillips, S.~E. Thorsett, \& S.~R. Kulkarni, Astronomical Society
  of the Pacific Conference Series, 19

\bibitem[Bauch 1990]{bau90}
Bauch, A. 1990, in { Impact of Pulsar Timing on Relativity and Cosmology}, ed.\
  D.~C. Backer, (Berkeley: Center for Particle Astrophysics), b1

\bibitem[Bell \& Bailes 1996]{bb96}
Bell, J.~F. \& Bailes, M. 1996, ApJ, 456, L33

\bibitem[Bell, Bailes, \& Bessell 1993]{bbb93}
Bell, J.~F., Bailes, M., \& Bessell, M.~S. 1993, Nature, 364, 603

\bibitem[Bell {et al.}  1996]{bbm+96}
Bell, J.~F., Bailes, M., Manchester, R.~N., Lyne, A.~G., Camilo, F., \& Sandhu,
  J.~S. 1996, MNRAS.
\newblock Submitted

\bibitem[Bell {et al.}  1995a]{bbm+95}
Bell, J.~F., Bailes, M., Manchester, R.~N., Weisberg, J.~M., \& Lyne, A.~G.
  1995a, ApJ, 440, L81

\bibitem[Bell \& Damour 1996]{bd96}
Bell, J.~F. \& Damour, T. 1996, Classical Quantum Gravity.
\newblock In Press

\bibitem[Bell {et al.}  1995b]{bkb+95}
Bell, J.~F., Kulkarni, S.~R., Bailes, M., Leitch, E.~M., \& Lyne, A.~G. 1995b,
  ApJ, 452, L121

\bibitem[Bignami \& Caraveo 1992]{bc92}
Bignami, G.~F. \& Caraveo, P.~A. 1992, Nature, 357, 287

\bibitem[Biraud {et al.}  1989]{bab+89}
Biraud, F., Aubry, D., Bougois, G., Darchy, B., Drouhin, J.~P., Lestrade,
  J.-F., \& Cognard, I. 1989, Unpublished

\bibitem[Boyd {et al.}  1995]{bcd+95}
Boyd, P.~T. {et al.}  1995, ApJ, 448, 365

\bibitem[Camilo, Foster, \& Wolszczan 1994]{cfw94}
Camilo, F., Foster, R.~S., \& Wolszczan, A. 1994, ApJ, 437, L39

\bibitem[Camilo {et al.}  1997a]{cmb+97}
Camilo, F., Manchester, R.~N., Bailes, M., Lyne, A.~G., \& Bell, J.~F. 1997a,
  ApJ.
\newblock In preparation

\bibitem[Camilo {et al.}  1997b]{cnk+97}
Camilo, F., Nice, D.~J., Kramer, M., Xilouris, K.~M., Sayer, R.~W., Shrauner,
  J.~A., \& Taylor, J.~H. 1997b, ApJ.
\newblock In preparation

\bibitem[Camilo, Thorsett, \& Kulkarni 1994]{ctk94}
Camilo, F., Thorsett, S.~E., \& Kulkarni, S.~R. 1994, ApJ, 421, L15

\bibitem[Clifton \& Lyne 1986]{cl86}
Clifton, T.~R. \& Lyne, A.~G. 1986, Nature, 320, 43

\bibitem[Cognard {et al.}  1993]{cbl+93}
Cognard, I., Bourgois, G., Lestrade, J.-F., Biraud, F., Aubry, D., Darchy, B.,
  \& Drouhin, J.-P. 1993, Nature, 366, 320

\bibitem[Cognard {et al.}  1995]{cbl+95}
Cognard, I., Bourgois, G., Lestrade, J.~F., Biraud, F., Aubry, D., Darchy, B.,
  \& Drouhin, J.~P. 1995, A\&A, 296, 169

\bibitem[Cook, Shapiro, \& Teukolsky 1994]{cst94}
Cook, G.~B., Shapiro, S.~L., \& Teukolsky, S.~A. 1994, ApJ, 423, L117

\bibitem[Cordes 1980]{cor80}
Cordes, J.~M. 1980, ApJ, 237, 216

\bibitem[Cordes, Pidwerbetsky, \& Lovelace 1986]{cpl86}
Cordes, J.~M., Pidwerbetsky, A., \& Lovelace, R. V.~E. 1986, ApJ, 310, 737

\bibitem[Damour \& Esposito-Far\`ese 1992]{de92a}
Damour, T. \& Esposito-Far\`ese, G. 1992, Phys. Rev. D, 46, 4128

\bibitem[Damour \& Sch\"afer 1991]{ds91}
Damour, T. \& Sch\"afer, G. 1991, Phys. Rev. Lett., 66, 2549

\bibitem[Damour \& Taylor 1991]{dt91}
Damour, T. \& Taylor, J.~H. 1991, ApJ, 366, 501

\bibitem[Damour \& Taylor 1992]{dt92}
Damour, T. \& Taylor, J.~H. 1992, Phys. Rev. D, 45, 1840

\bibitem[Doroshenko \& Kopeikin 1995]{dk95}
Doroshenko, O.~V. \& Kopeikin, S.~M. 1995, MNRAS, 274, 1029

\bibitem[Gouiffes, Finley, \& \"{O}gelman 1992]{gfo92}
Gouiffes, C., Finley, J.~P., \& \"{O}gelman, H. 1992, ApJ, 394, 581

\bibitem[Grenier {et al.}  1994]{gbb+94}
Grenier, I. A.;~Bennett, K., Bucheri, R., Gros, M., Henriksen, R.~N., Hermsen,
  W., Kanbach, G., \& Sacco, B. 1994, ApJ, 276, 813

\bibitem[Groth 1975]{gro75c}
Groth, E.~J. 1975, ApJS, 29, 453

\bibitem[Gupta 1995]{gup95}
Gupta, Y. 1995, ApJ, 451, 717

\bibitem[Hamilton \& Lyne 1988]{hl88}
Hamilton, P.~A. \& Lyne, A.~G. 1988, MNRAS, 224, 1073

\bibitem[Hankins 1971]{han71}
Hankins, T.~H. 1971, ApJ, 169, 487

\bibitem[Hankins, Stinebring, \& Rawley 1987]{hsr87}
Hankins, T.~H., Stinebring, D.~R., \& Rawley, L.~A. 1987, ApJ, 315, 149

\bibitem[Hogan \& Rees 1984]{hr84}
Hogan, C.~J. \& Rees, M.~J. 1984, Nature, 311, 109

\bibitem[Hulse \& Taylor 1974]{ht74}
Hulse, R.~A. \& Taylor, J.~H. 1974, ApJ, 191, L59

\bibitem[Johnston 1996]{joh96}
Johnston, D. 1996, in { Pulsars: Problems and Progress, {IAU} Colloquium 160},
  ed.\ S. Johnston, M.~A. Walker, \& M. Bailes, Astronomical Society of the
  Pacific, 501

\bibitem[Johnston {et al.}  1992]{jlm+92}
Johnston, S., Lyne, A.~G., Manchester, R.~N., Kniffen, D.~A., D'Amico, N., Lim,
  J., \& Ashworth, M. 1992, MNRAS, 255, 401

\bibitem[Johnston {et al.}  1996]{jml+96}
Johnston, S., Manchester, R.~N., Lyne, A.~G., D'Amico, N., Bailes, M.,
  Gaensler, B.~M., \& Nicastro, L. 1996, MNRAS, 279, 1026

\bibitem[Kaspi 1995]{kas95}
Kaspi, V.~M. 1995, in { Millisecond Pulsars - A Decade of Surprise}, ed.\ A.~S.
  Fruchter, M. Tavani, \& D.~C. Backer, volume~72, Astronomical Society of the
  Pacific, 345

\bibitem[Kaspi 1996]{khnt96}
Kaspi, V.~M. 1996, in { 31st Scientific Assembly of COSPAR: E1.5 Satellite and
  Ground Based Studies of Radiopulsars}, ed.\ P.~A. Caraveo, Elsevier Science,
  Ltd., In Press

\bibitem[Kaspi {et al.}  1996]{kbm+96}
Kaspi, V.~M., Bailes, M., Manchester, R.~N., Stappers, B.~W., \& Bell, J.~F.
  1996, Nature, 381, 584

\bibitem[Kaspi, Tauris, \& Manchester 1996]{ktm96}
Kaspi, V.~M., Tauris, T., \& Manchester, R.~N. 1996, ApJ, 459, 717

\bibitem[Kaspi {et al.}  1995]{ktn+95}
Kaspi, V.~M., Tavani, M., Nagase, F., Hirayama, M., Hoshino, M., Aoki, T.,
  Kawai, N., \& Arons, J. 1995, ApJ, 453, 424

\bibitem[Kaspi, Taylor, \& Ryba 1994]{ktr94}
Kaspi, V.~M., Taylor, J.~H., \& Ryba, M. 1994, ApJ, 428, 713

\bibitem[Kopeikin 1995]{kop95}
Kopeikin, S.~M. 1995, ApJ, 439, L5

\bibitem[Kopeikin 1996]{kop96}
Kopeikin, S.~M. 1996, ApJ, 467, L93

\bibitem[Kulkarni 1986]{kul86}
Kulkarni, S.~R. 1986, ApJ, 306, L85

\bibitem[Kulkarni 1992]{kul92}
Kulkarni, S.~R. 1992, Philos. Trans. Roy. Soc. London A, 341, 77

\bibitem[Lai, Bildsten, \& Kaspi 1995]{lbk95}
Lai, D., Bildsten, L., \& Kaspi, V.~M. 1995, ApJ, 452, 819

\bibitem[Lohsen 1975]{loh75}
Lohsen, E. 1975, Nature, 258, 688

\bibitem[Lohsen 1981]{loh81}
Lohsen, E. H.~G. 1981, A\&AS, 44, 1

\bibitem[Lundgren, Camilo, \& Foster 1996]{lfc96}
Lundgren, S.~C., Camilo, F., \& Foster, R.~S. 1996, in { Pulsars: Problems and
  Progress, {IAU} Colloqium 160}, ed.\ S. Johnston, M.~A. Walker, \& M. Bailes,
  Astronomical Society of the Pacific, 497

\bibitem[Lundgren {et al.}  1995]{lcu+95}
Lundgren, S.~C., Cordes, J.~M., Ulmer, M., Matz, S.~M., Lomatch, S., Foster,
  R.~S., \& Hankins, T. 1995, ApJ, 453, 433

\bibitem[Lyne {et al.}  1996a]{lkb+96}
Lyne, A.~G., Kaspi, V.~M., Bailes, M., Manchester, R.~N., Taylor, H., \&
  Arzoumanian, Z. 1996a, MNRAS, 281, L14

\bibitem[Lyne \& Lorimer 1994]{ll94}
Lyne, A.~G. \& Lorimer, D.~R. 1994, Nature, 369, 127

\bibitem[Lyne, Manchester, \& Taylor 1985]{lmt85}
Lyne, A.~G., Manchester, R.~N., \& Taylor, J.~H. 1985, MNRAS, 213, 613

\bibitem[Lyne {et al.}  1996b]{lpgc96}
Lyne, A.~G., Pritchard, R.~S., Graham-Smith, F., \& Camilo, F. 1996b, Nature,
  381, 497

\bibitem[Lyne, Pritchard, \& Smith 1993]{lps93}
Lyne, A.~G., Pritchard, R.~S., \& Smith, F.~G. 1993, MNRAS, 265, 1003

\bibitem[Lyne \& Smith 1990]{ls90}
Lyne, A.~G. \& Smith, F.~G. 1990, { Pulsar Astronomy}, (: Cambridge University
  Press)

\bibitem[Manchester {et al.}  1993]{mml+93}
Manchester, R.~N., Mar, D., Lyne, A.~G., Kaspi, V.~M., \& Johnston, S. 1993,
  ApJ, 403, L29

\bibitem[Manchester \& Peterson 1989]{mp89}
Manchester, R.~N. \& Peterson, B.~A. 1989, ApJ, 342, L23

\bibitem[Manchester, Taylor, \& Van 1974]{mtv74}
Manchester, R.~N., Taylor, J.~H., \& Van, Y.-Y. 1974, ApJ, 189, L119

\bibitem[Manchester {et al.}  1980]{mwpe80}
Manchester, R.~N., Wallace, P.~T., Peterson, B.~A., \& Elliott, K.~H. 1980,
  MNRAS, 190, 9P

\bibitem[Martin, Lyne, \& Pritchard 1997]{mlp97}
Martin, C.~E., Lyne, A.~G., \& Pritchard, R.~S. 1997, MNRAS.
\newblock In preparation

\bibitem[Mattox {et al.}  1992]{mbf92}
Mattox, J.~R., Bertsch, D.~L., Fichtel, C.~E., Hatman, R.~C., Kniffen, D.~A.,
  \& Thompson, D.~J. 1992, ApJ, 401, L23

\bibitem[McCulloch 1996]{mcc96}
McCulloch, P.~M. 1996, in { Pulsars: Problems and Progress, {IAU} Colloquium
  160}, ed.\ S. Johnston, M.~A. Walker, \& M. Bailes, Astronomical Society of
  the Pacific, 97

\bibitem[McCulloch, Taylor, \& Weisberg 1979]{mtw79}
McCulloch, P.~M., Taylor, J.~H., \& Weisberg, J.~M. 1979, ApJ, 227, L133

\bibitem[Middleditch, Pennypacker, \& Burns 1987]{mpb87}
Middleditch, J., Pennypacker, C.~R., \& Burns, M.~S. 1987, ApJ, 315, 142

\bibitem[Nagase {et al.}  1990]{ndl+90}
Nagase, F., Deeter, J., Lewis, W., Dotani, T., Makino, F., \& Mitsuda, K. 1990,
  ApJ, 351, L13

\bibitem[Navarro 1994]{nav94}
Navarro, J. 1994.
\newblock PhD thesis, California Institute of Technology

\bibitem[Nice, Sayer, \& Taylor 1996]{nst96}
Nice, D.~J., Sayer, R.~W., \& Taylor, J.~H. 1996, ApJ, 466, L87

\bibitem[Nice \& Taylor 1995]{nt95}
Nice, D.~J. \& Taylor, J.~H. 1995, ApJ, 441, 429

\bibitem[Nice \& Thorsett 1996]{nt96}
Nice, D.~J. \& Thorsett, S.~E. 1996, in { Pulsars: Problems and Progress, {IAU}
  Colloquium 160}, ed.\ S. Johnston, M.~A. Walker, \& M. Bailes, Astronomical
  Society of the Pacific, 523

\bibitem[Phinney 1992]{phi92b}
Phinney, E.~S. 1992, Philos. Trans. Roy. Soc. London A, 341, 39

\bibitem[Phinney \& Kulkarni 1994]{pk94}
Phinney, E.~S. \& Kulkarni, S.~R. 1994, ARAA, 32, 591

\bibitem[Press {et al.}  1992]{ptvf92}
Press, W.~H., Teukolsky, S.~A., Vetterling, W.~T., \& Flannery, B.~P. 1992, {
  Numerical Recipes: {T}he Art of Scientific Computing, 2$^{nd}$ edition},
  (Cambridge: Cambridge University Press)

\bibitem[Rickett 1990]{ric90}
Rickett, B.~J. 1990, ARAA, 28, 561

\bibitem[Romani \& Taylor 1983]{rt83}
Romani, R.~W. \& Taylor, J.~H. 1983, ApJ, 265, L35

\bibitem[Ruderman 1969]{rud69}
Ruderman, M. 1969, Nature, 223, 597

\bibitem[Ryba \& Taylor 1991]{rt91a}
Ryba, M.~F. \& Taylor, J.~H. 1991, ApJ, 371, 739

\bibitem[Sandhu {et al.}  1996]{sbm+96}
Sandhu, J.~S., Bailes, M., Manchester, R.~N., Navarro, J., Kulkarni, S.~R., \&
  Anderson, S.~B. 1996, ApJ.
\newblock in preparation

\bibitem[Schneider 1989]{sch89b}
Schneider, J. 1989, A\&A, 214, 1

\bibitem[Schneider 1990]{sch90}
Schneider, J. 1990, A\&A, 232, 62

\bibitem[Seward, Harnden, \& Helfand 1984]{shh84}
Seward, F.~D., Harnden, F.~R., \& Helfand, D.~J. 1984, ApJ, 287, L19

\bibitem[Shemar 1995]{she95a}
Shemar, S.~L. 1995.
\newblock PhD thesis, University of Manchester

\bibitem[Shemar \& Lyne 1996]{sl96}
Shemar, S.~L. \& Lyne, A.~G. 1996, MNRAS.
\newblock In press

\bibitem[Shemar, Lyne, \& Smith 1996]{sls96}
Shemar, S.~L., Lyne, A.~G., \& Smith, F.~G. 1996, MNRAS.
\newblock In prep.

\bibitem[Shklovskii 1970]{shk70}
Shklovskii, I.~S. 1970, Sov. Astron., 13, 562

\bibitem[Standish 1990]{sta90}
Standish, E.~M. 1990, A\&A, 233, 252

\bibitem[Stappers {et al.}  1996]{sbl+96}
Stappers, B.~W. {et al.}  1996, ApJ, 465, L119

\bibitem[Tavani, Arons, \& Kaspi 1994]{tak94}
Tavani, M., Arons, J., \& Kaspi, V.~M. 1994, ApJ, 433, L37

\bibitem[Taylor 1992]{tay92}
Taylor, J.~H. 1992, Philos. Trans. Roy. Soc. London A, 341, 117

\bibitem[Taylor 1996]{tay96}
Taylor, J.~H. 1996, in { Pulsars: Problems and Progress, {IAU} Colloquium 160},
  ed.\ S. Johnston, M.~A. Walker, \& M. Bailes, Astronomical Society of the
  Pacific, 65

\bibitem[Taylor \& Cordes 1993]{tc93}
Taylor, J.~H. \& Cordes, J.~M. 1993, ApJ, 411, 674

\bibitem[Taylor \& Weisberg 1989]{tw89}
Taylor, J.~H. \& Weisberg, J.~M. 1989, ApJ, 345, 434

\bibitem[Taylor {et al.}  1992]{twdw92}
Taylor, J.~H., Wolszczan, A., Damour, T., \& Weisberg, J.~M. 1992, Nature, 355,
  132

\bibitem[Thompson 1993]{tho93}
Thompson, D.~J. 1993, in { Isolated Pulsars}, ed.\ K.~A. van Riper, R. Epstein,
  \& C. Ho, (New York: Cambridge University Press), 385

\bibitem[Thorsett 1996]{tho96}
Thorsett, S.~E. 1996, Phys. Rev. Lett., 77, 1432

\bibitem[Thorsett \& Dewey 1996]{td96}
Thorsett, S.~E. \& Dewey, R.~J. 1996, Phys. Rev. D, 53, 3468

\bibitem[Thorsett \& Phillips 1992]{tp92}
Thorsett, S.~E. \& Phillips, J.~A. 1992, ApJ, 387, L69

\bibitem[van Kerkwijk 1996]{van96}
van Kerkwijk, M.~H. 1996, in { Pulsars: Problems and Progress, {IAU} Colloquium
  160}, ed.\ S. Johnston, M.~A. Walker, \& M. Bailes, Astronomical Society of
  the Pacific, 489

\bibitem[van Kerkwijk, Bergeron, \& Kulkarni 1996]{vbk96}
van Kerkwijk, M.~H., Bergeron, P., \& Kulkarni, S.~R. 1996, ApJ, 467, L89

\bibitem[van Paradijs 1996]{vp96}
van Paradijs, J. 1996, in { {Pulsar Timing, General Relativity and the Internal
  Structure of Neutron Stars}}, ed.\ E. van~den Heuvel, J. van Paradijs, \& Z.
  Arzoumanian, Elsevier Science, In preparation

\bibitem[Wex 1996]{wex96}
Wex, N. 1996, A\&A.
\newblock In Press

\bibitem[Wolszczan 1994]{wol94}
Wolszczan, A. 1994, Science, 264, 538

\bibitem[Wolszczan \& Frail 1992]{wf92}
Wolszczan, A. \& Frail, D.~A. 1992, Nature, 355, 145

\end{thebibliography}

\end{document}